\documentclass[aps,prc,superscriptaddress,nofootinbib,floatfix,preprint,tightenlines]{revtex4-1}
\usepackage{bbm}
\usepackage{amsthm}
\usepackage{mathrsfs}
\usepackage{graphicx}
\usepackage{float} 
\usepackage{amsmath}
\usepackage{subfigure}
\usepackage{multirow}
\usepackage{amsfonts}
\usepackage[dvipsnames]{xcolor}
\newcommand{\eq}{\begin{eqnarray}}
\newcommand{\en}{\end{eqnarray}}

\begin{document}

\title{An alternative scheme for 
pionless EFT: neutron-deuteron scattering in the doublet $S$-wave}

\author{M.~Ebert}
\email{mebert@theorie.ikp.physik.tu-darmstadt.de}
\affiliation{Technische Universit\"at Darmstadt, Department of Physics,
64289 Darmstadt, Germany}

\author{H.-W.~Hammer}
\email{hans-werner.hammer@physik.tu-darmstadt.de}
\affiliation{Technische Universit\"at Darmstadt, Department of Physics,
64289 Darmstadt, Germany}
\affiliation{ExtreMe Matter Institute EMMI and Helmholtz Forschungsakademie
  Hessen f\"ur FAIR (HFHF),
64291 Darmstadt, Germany}

\author{A.~Rusetsky}
\email{rusetsky@hiskp.uni-bonn.de}
\affiliation{HISKP and BCTP, Rheinische Friedrich-Wilhelms Universit\"at Bonn, 53115 Bonn, Germany}
\affiliation{Tbilisi State  University,  0186 Tbilisi, Georgia\vspace*{.3cm}}

\vspace*{.3cm}

\begin{abstract}
Using the effective-range expansion for the two-body amplitudes may generate spurious sub-threshold poles 
outside of the convergence range of the expansion. In the
infinite volume, the emergence of such poles leads to the breakdown of unitarity in the three-body amplitude.
We discuss the extension of our alternative subtraction scheme for including  effective range corrections in pionless effective field theory for spinless bosons to nucleons. In particular, we consider the neutron-deuteron system in the doublet $S$-wave channel explicitly. 
\end{abstract}

\maketitle

\section{Introduction}
Pionless EFT describes the physics of nucleons at very 
low-energy in terms of contact interactions between the 
nucleons~\cite{vanKolck:1997ut,Kaplan:1998tg,Kaplan:1998we,vanKolck:1998bw,Chen:1999tn}. It can be understood as an expansion around
the interacting renormalization group fixed point associated with
the unitary limit of infinite scattering
length. The breakdown scale of pionless EFT is set by the
pion mass, $M_{high} \sim M_\pi$, while the typical low-energy scale is determined the scattering length $a$ and
the involved momenta $k$, $M_{low}\sim 1/a \sim k$. For momenta
of order of the breakdown scale,
$k\sim M_\pi$, pion exchange can no longer be treated as a short-range interaction and has to be included explicitly.
For reviews of EFT's of nuclear forces at different resolution scales we refer the reader to
Refs.~\cite{Bedaque:2002mn,Epelbaum:2008ga,Hammer:2019poc}.

This EFT is also very efficient for the formulation of the three-body problem in a finite volume that is motivated by the necessity of analyzing lattice data in the three-particle sector. Namely, one of three existing formulations of the three-body
quantization condition (an equation that connects the finite-volume energy spectrum with
the infinite-volume observables in the three-particle system), the so-called Nonrelativistic Effective Field Theory (NREFT) approach~\cite{Hammer:2017uqm, Hammer:2017kms}, is directly based on it. A Lorentz-invariant extension of the NREFT approach is possible and was presented in Ref.~\cite{Muller:2021uur}. Two other conceptually equivalent formulations are  
the so-called Relativistic Field Theory (RFT)~\cite{Hansen:2014eka, Hansen:2015zga} and Finite Volume Unitarity (FVU)~\cite{Mai:2017bge,Mai:2018djl} approaches. For a more detailed overview of the rapid progress in this field,
we refer the reader to the two recent reviews on the subject~\cite{Hansen:2019nir,Mai:2021lwb}.

At leading order (LO) in pionless EFT, one needs to resum a 
momentum-independent contact interaction in order to describe the large
scattering length physics~\cite{vanKolck:1997ut,Kaplan:1998tg,Kaplan:1998we,vanKolck:1998bw,Chen:1999tn}. This resummation is conveniently
implemented using dibaryon or dimer fields.
At next-to-leading order (NLO) the two-body ranges enter perturbatively~\cite{Hammer:2001gh,Bedaque:1998km,Ji:2011qg,Ji:2012nj,Vanasse:2013sda}. At higher orders, the procedure of perturbative range insertions becomes tedious. 

Moreover, in a finite volume the perturbative approach
becomes problematic since integrals are replaced by discrete sums which will not converge near the singularities of
the dimer propagator.
Specifically, in a finite box of size $L$
the dimer propagator gets replaced by~\cite{Hammer:2017uqm,Hammer:2017kms}:
\eq\label{eq:tauL}
D_L({\bf k},{k^*}^2)=\frac{1}{k^*\cot\delta(k^*)+S({\bf k},{k^*}^2)}\, .
\en
Here, ${\bf k},k^*$ denote the total three-momentum of a dimer and
the magnitude of the relative momentum of two particles, constituting
a dimer, in their center-of-mass frame. Furthermore, $\delta(k^*)$ is
the phase shift and the quantity $S({\bf k},{k^*}^2)$ stands for
the infinite sum
\eq\label{eq:S}
S({\bf k},{k^*}^2)=-\frac{4\pi}{L^3}\,\sum_{\bf p}\frac{1}{{\bf p}^2+{\bf p}{\bf k}+{\bf k}^2-mE}\, ,\quad\quad {\bf p}=\frac{2\pi}{L}\,{\bf n}\, ,\quad
{\bf n}=\mathbb{Z}^3\, ,
\en
where $E$ is the total energy of the particle-dimer system in the rest frame.
Thus the propagator has an infinite tower of poles above the elastic threshold,
corresponding to the finite-volume energy spectrum in the two-particle
subsystems. The perturbative expansion will not work in the vicinity of these
poles, producing denominators that are more and more singular.
In the infinite volume,
these poles condense and form an elastic cut.
The sum turns into an integral that can be easily evaluated, leading to a well-known result.

To avoid this problem, range corrections can be resummed by including the effective range in the denominator of the dimer propagator. It is well known that such a resummation introduces a spurious pole in the deuteron propagator at
a momentum scale of roughly 200 MeV \cite{Bedaque:1997qi}.  Since it is outside the range of validity of the EFT it can be considered an irrelevant UV artifact. However, in three- and higher-body systems it can limit the range of cutoffs that can be used in the numerical solution of the scattering equations. In the three-nucleon system, this is especially true in the doublet S-wave of neutron-deuteron scattering (triton channel) unless measures are taken to remove the pole. In the quartet S-wave,
due to the Pauli principle, the solution is not sensitive to this deep pole and the cutoff can be made arbitrarily large \cite{Bedaque:1997qi,Bedaque:1998mb}.

In Ref.~\cite{Ebert:2021epn}, we have developed a 
method to remove the spurious poles for the simpler case
of spinless bosons and in Ref.~\cite{Pang:2022nim}
this method was tested in finite volume calculations.
Here, we extend this solution 
to doublet S-wave of neutron-deuteron scattering (i.e., the $^2S_{1/2}$-channel)
and the triton, which is relevant for lattice QCD calculations of the three-nucleon system.

\section{Integral equations for the doublet channel}

Using the dimer formalism, the neutron-deuteron scattering process in the $^2S_{1/2}$-channel is determined up to next-to-next-to-leading order
(N$^2$LO) in pionless EFT by 
the two amplitudes $T_t$ and $T_s$. The amplitude $T_t$
describes the elastic scattering of a dimer field for the deuteron ($d_t$) and a neutron ($n$), 
$d_t+n\rightarrow d_t+n$, 
while $T_s$ describes the scattering of the same 
initial state into a spin-singlet 
dimer field ($d_s$) and a neutron, $d_t+N\rightarrow d_s+N$.
The amplitudes are given as solutions of the following coupled Faddeev equations (cf.~Eq.~(29) in \cite{Bedaque:2002yg}):
\begin{align}
  \begin{split}
    T_s(p,k,E)=&\frac{1}{4}\bigg(3K(p,k,E)+2H(E,\Lambda)\bigg)\\
    &+\frac{1}{2\pi}\int_0^\Lambda dq\, q^2 \bigg{[}\bigg( K(p,q,E) +2H(E,\Lambda)\bigg){D}_s(q,E)T_s(q,k,E)\\
    &\qquad+\bigg(3K(p,q,E)+2H(E,\Lambda)\bigg){D}_t(q,E)T_t(q,k)
    \bigg{]},\\
    T_t(p,k,E)=&\frac{1}{4}\bigg(K(p,k,E)+2H(E,\Lambda)\bigg)\\
    &+\frac{1}{2\pi}\int_0^\Lambda dq\, q^2 \bigg{[}\bigg(K(p,q,E) +2H(E,\Lambda)\bigg){D}_t(q,E)T_t(q,k,E)\\
    &\qquad+\bigg(3K(p,q,E)+2H(E,\Lambda)\bigg){D}_s(q,E)T_s(q,k,E)\bigg{]},
    \label{STMeq}
 \end{split}
\end{align}
where $E$ is the total energy and $p$, $k$ are the relative momenta of the neutron and the dimer in the incoming and outgoing channel, respectively.
Note that only $T_t$ corresponds to a physical scattering process and $k$ can be set on shell, such
that the total energy is  $E=3k^2/(4m)-\gamma^2/m$ with $\gamma$ the deuteron binding momentum and $m$ the nucleon mass.
The interaction is given by the $S$-wave projected  one-particle exchange, 
\begin{align}
    K(p,q,E)=\frac{1}{pq}\ln\left(\frac{p^2+pq+q^2-mE}{p^2-pq+q^2-mE}\right),
    \label{Exchange}
\end{align}
and a series of three-body contact interactions
\begin{align}
H(E,\Lambda)=\frac{2H_0(\Lambda)}{\Lambda^2}+\frac{2H_2(\Lambda)}{\Lambda^4}(mE+\gamma^2)+\cdots.
\end{align}
According to the power-counting of pionless EFT at LO and NLO only the term proportional to $H_0$ contributes. At N$^2$LO also the term proportional to $H_2$ contributes. If the effective range corrections are included non-perturbatively by resumming all diagramms, the effective range $r_{t/s}$ in the spin triplet/singlet channels of the nucleon nucleon interaction
appears in the denominator of the respective dimer propagators
\begin{align}
     { D}_{t/s}(q,E)=\frac{1}{-1/a_{t/s}-\frac{r_{t/s}}{2}q^{*2}+q^*},
     \label{eq:dimerprop}
\end{align}
where $q^*$ is the magnitude of the boosted relative momentum, i.e.,
\eq
\label{eq:defq}
{q^*}^2=\frac{3}{4}\,q^2-mE\,.
\en
The effective range parameters in the spin-triplet channel are
$a_t=5.42\text{ fm}$ and $r_t= 1.76\text{ fm}$. These values imply a deuteron binding energy of 2.2 MeV. The corresponding low-energy parameters in the spin-singlet channel are  $a_s=-23.71\text{ fm}$ and  $r_s=2.37\text{ fm}$. As discussed above, the propagator \eqref{eq:dimerprop} exhibits two poles per channel:
\begin{align}
    q^*_{1,t/s}= \frac{2/a_{t/s}}{1+\sqrt{1-2r_{t/s}/a_{t/s}}}\simeq \frac{1}{a_{t/s}};  \quad q^*_{2,t/s}=\frac{1+\sqrt{1-2r_{t/s}/a_{t/s}}}{r_{t/s}} \simeq \frac{2}{r_{t/s}}.
\end{align}
The first one is a low-momentum pole and has physical significance. In the triplet channel this corresponds to the bound state pole of the deuteron at $q^*_{1,t}=45.72\text{ MeV}$. In the singlet channel there is a virtual state at $q^*_{2,s}=-7.89\text{ MeV}$.
The second pole is spurious. It is an artefact of the non-perturbative resummation of the effective range corrections and located at high momenta of order $1/r_{t/s}$ where pion degrees of freedom become important. In principle one could simply discard this pole. However, it lies on the path of integration in Eq.~(\ref{STMeq}) for $q^*_{2,t}= 178.50 \text{ MeV}$ as well as for $q^*_{2,s}=152.45 \text{ MeV}$. In the case of the spin-triplet dimer it has a negative residue and leads violation of unitarity. To
avoid this problem, while still including the effective range non-perturbatively,  
we proposed in Ref.~\cite{Ebert:2021epn} to replace the propagator \eqref{eq:dimerprop}
by a modified propagator ${\bar {D}}_{t/s}(q,E)$ where the spurious pole is subtracted. In our case, the proposed scheme amounts to
using
\begin{align}
\label{eq:exp}
{\bar { D}}_{t/s}(q,E)=\frac{2 (q^{*}_{1,t/s}+q^{*}_{2,t/s})/r_{t/s}}{ (q^{*}_{2,t/s}-q^{*}_{1,t/s})(q^{*}+q^{*}_{2,t/s})}\frac{1}{q^{*}-q^{*}_{1,t/s}}+\frac{4 q^*_{2,t/s}/r_{t/s}}{ (q^*_{2,t/s}-q^*_{1,t/s})\,q^{*2}_{2,t/s}}\left(1+\frac{q^{*2}}{q^{*2}_{2,t/s}}+\cdots\right)
\end{align}
instead of  ${\cal D}_{t/s}(q,E)$ in Eq.~(\ref{STMeq}).
The ellipses stand for higher order terms of the expansion of the spurious contribution. At LO in the power counting, $r_{t/s}=0$ and the spurious pole is absent. At NLO we use only the constant part of the expansion of the last term in Eq. (\ref{eq:exp}), while both 
the constant and quadratic term in $q^*$ are included at N$^2$LO.
At each higher order there is a finite renormalization of the three-body force $H(E,\Lambda)$. The justification for this procedure will be discussed in the next section. For more technical details we refer the reader to Ref.~\cite{Ebert:2021epn}.

\section{$SU(4)$ symmetry and choice of subtractions}
\label{sec:su4}

In our previous paper~\cite{Ebert:2021epn} it was demonstrated that the
subtraction of the spurious pole in the dimer propagator,  Eqs.~(\ref{eq:dimerprop}) and (\ref{eq:exp}), at a given
order in the low-energy expansion, can be compensated
by adding a low-energy polynomial to the driving term of the Faddeev equation
for the particle-dimer scattering. Consequently,
the subtraction was interpreted 
merely as a change of the renormalization prescription in the particle-dimer
Lagrangian. The method contained a caveat, however, since the coefficients
of this polynomial turned out to be complex. It was argued that this fact
reflects the unphysical nature of the spurious poles, and dropping of a complex
polynomial was interpreted as a part of rectifying the formalism through removal of
the spurious pole. As a consequence, at each higher order there is a finite renormalization of the three-body force $H(E,\Lambda)$
but no further modifications arise.
Thus it would be fair to say that the ``subtracted'' form
of the Faddeev equations has been established by applying very plausible but still intuitive arguments to the ``unsubtracted'' equation containing the spurious pole.

The situation turns out to be more subtle in case of three nucleons. In the $^2S_{1/2}$ channel of neutron-deuteron scattering, one has to deal with two dimers, $d_s$ and $d_t$, as discussed above. 
The Faddeev equations for this case, Eq. (\ref{STMeq}),
are written down assuming the $SU(4)$ symmetry of the three-nucleon Lagrangian
and thus contain only {\em one} three-body force ${ H}(E,\Lambda)$ \cite{Bedaque:2002yg}.
However, at
LO in the low-energy expansion, the $SU(4)$ symmetry in the
three-particle sector is
not an assumption, since only one term with six fermion fields and
no derivatives can be written down -- all other terms can be reduced to it by
using Fierz transformations.

In order to ease the further discussion, let us assume that the exact
$SU(4)$ symmetry holds in the two-particle sector, i.e.,
${D}_{s}(q)={ D}_{t}(q)\equiv{ D}(q)$. Furthermore, we assume that there are only non-derivative particle-dimer interactions. As shown in Ref.~\cite{Bedaque:2002yg},
in this case it is convenient to work with the amplitudes $T_\pm=T_s\pm T_t$. The Faddeev equations for these amplitudes decouple, and the equation for $T_-$ does not contain the polynomial three-body term ${ H}(E,\Lambda)$ at all: 
\eq
T_-(p,k,E)=\frac{1}{2}\,{ K}(p,k,E)
-\frac{1}{\pi}\,\int_0^\Lambda dq\, q^2 {K}(p,q,E){ D}(q,E)T_-(q,k,E)\, .
\en
At this point, we may try to proceed in the same way as in the case of three
identical bosons~\cite{Ebert:2021epn}. In order to subtract the spurious
pole, we may write:
\eq
{D}(q)=\bar{D}(q)+f(q)\, ,
\en
where the propagator $\bar{D}(q)$ does not contain the spurious pole
anymore, and $f(q)$ is a low-energy polynomial in the variable ${q^*}^2$ defined in Eq. (\ref{eq:defq}).
The original Faddeev equation can then be split into two equations:
\eq
\label{eq:split}
T_-(p,k,E)&=&\frac{1}{2}\,{ W}(p,k,E)
-\frac{1}{\pi}\,\int_0^\Lambda dq \,q^2 {W}(p,q,E)\bar {D}(q)T_-(q,k,E)\, ,
\nonumber\\[2mm]
{W}(p,k,E)&=&{K}(p,k,E)
-\frac{1}{\pi}\,\int_0^\Lambda dq \,q^2 {K}(p,q,E)f(q){W}(q,k,E)\, .
\en
A problem that one is facing is immediately seen from the above equations.
Namely, the new kernel ${ W}$ contains a low-energy polynomial that starts
at lowest-order in the EFT expansion (i.e., it does not vanish
when $p=k=0$ and $E=-\gamma^2/m$). On the other hand, such a term
cannot be removed by the renormalization, because we do not have a $SU(4)$
non-symmetric three-body term with no derivatives at our disposal.

In order to circumvent this problem, one has to look more carefully at its
roots. First of all, such a problem cannot arise, in principle, if one
faithfully applies perturbation theory in the EFT, using dimensional regularization,
and carries out
calculations of the amplitude at a given finite order. In this case, there
are neither spurious poles nor is $SU(4)$ violated. However, in our case
using strict perturbation theory
is not an option, and resummations in
the two-particle sector are necessary. The spurious pole(s) leading, in
particular, to the violation of three-particle unitarity already at low
energies emerge exactly as a result of this resummation. Hence, the subtraction
procedure is nothing but an attempt to remove
the unwanted artifacts of the resummation. 
Recall now that, by construction, the first $n$
terms of $f(q)$ in the expansion in ${q^*}^2$ vanish ($n$ denotes the order
of the subtraction). Hence, in the range of validity of the EFT expansion,
the quantity $f(q)$ can be made arbitrarily small. Then, choosing a cutoff $\Lambda$ smaller
than the position of the spurious pole (which is of order of the breakdown scale of the theory),
it is seen that the difference between ${W}(p,k,E)$ and ${K}(p,k,E)$ consequently
tends to zero, as the number of subtractions increases.

To summarize, it is clear that one may set ${ W}(p,k,E)={K}(p,k,E)$ in the effective
theory with a sufficiently low cutoff $\Lambda$. One may now {\em postulate} this property in the
subtracted Faddeev equations (irrespective of the value of cutoff),
arguing that this identification merely amounts to a
resummation in the two-body sector, which is different from the standard (``unsubtracted'')
form of resummation (leading to the emergence of the spurious poles).
Both resummations are equivalent at low energies and differ only at momenta
where the EFT expansion is no more valid. Hence, the ``subtracted'' form
of the Faddeev equations can be treated on equal footing with the standard
one. Thus we may proceed with the solution of the subtracted 
Faddeev equations discussed in the previous section.

\section{Results and discussion}

In the following we demonstrate that this strategy indeed works in the $^2S_{1/2}$ channel of neutron-deuteron scattering. We calculate the real and imaginary parts of the elastic scattering phase shift and determine
the LO three-body force, $H_0$, such that the experimental value of the $^2S_{1/2}$ neutron-deuteron scattering length,
$a_{nd}^{1/2} = (0.65\pm 0.04)$~fm \cite{Dilg:1971gqb}, is reproduced. The second three-body force $H_2$, which enters at N$^2$LO, is matched to the real part of the doublet scattering phase shift at finite energy. Here we use the value obtained
by Kievsky et al.~\cite{Kievsky:1996ca} with Argonne V 18 and an Urbana IX three-body force at finite momentum $p\approx 85$ MeV. There is also an
experimental phase shift analysis from the 1960's \cite{vanOers:1967lny}
that is consistent with the calculation of Kievsky et al.
However, there are large fluctuations in the extracted phase shift as a function of the energy such that it of limited use for our purpose. Therefore this phase shift analysis is not included in
our comparison.

\begin{figure}[H]
\center
\includegraphics[width=0.75\linewidth]{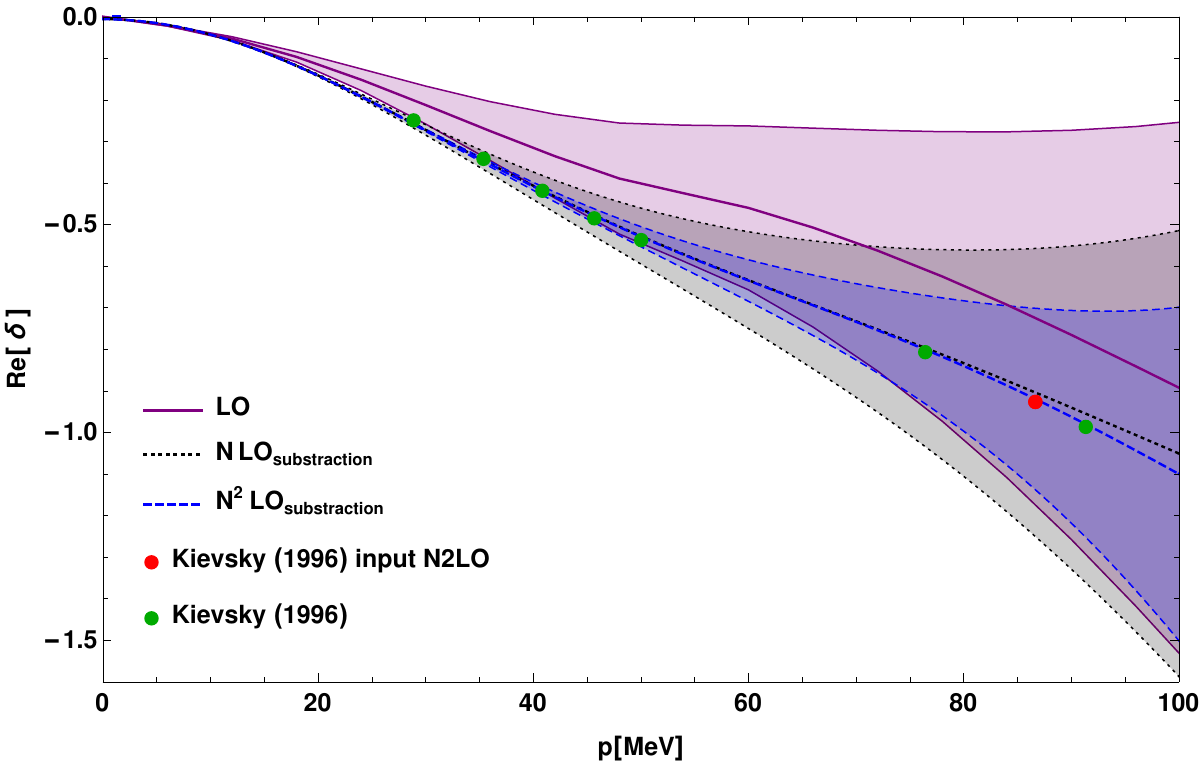}
\caption{The real part of the neutron-deuteron $^2S_{1/2}$ phase shift in our alternative subtraction scheme. The purple solid line is the result at LO, while the black dotted 
blue dashed lines give our result at NLO and N$^2$LO. respectively.
The theoretical uncertainties is given by the shaded bands.
For comparison, we show he calculation 
by Kievsky et al.~\cite{Kievsky:1996ca} (circles).
The data point indicated by the red circle was used as input to
match the N$^2$LO three-body force $H_2$.}
\label{PhaseshiftRe}
\end{figure}

In Fig.~\ref{PhaseshiftRe}, we compare our calculation of the 
real part of the neutron-deuteron $^2S_{1/2}$ phase shift with the calculation 
by Kievsky et al.~\cite{Kievsky:1996ca} (green circles). The red
circle indicates the input value from \cite{Kievsky:1996ca}
used to determine the N$^2$LO three-body force $H_2$.
The purple line is the result at LO. The black dotted line is the result at NLO with our subtraction scheme. The corresponding 
result at N$^2$LO is shown as the dashed blue curve.
The theoretical uncertainty in our calculation is estimated as $(p/\Lambda_b)^i$ with $i=1,2,3$ at LO, NLO, N$^2$LO, respectively. 
The breakdown scale $\Lambda_b=140\text{ MeV}$ is taken as the pion mass. The corresponding uncertainties are shown by the shaded bands.
The calculations at LO, NLO, N$^2$LO show the expected convergence behavior. They are consistent
with the their error estimates and the phenomenological calculation by Kievsky et al. \cite{Kievsky:1996ca} is well reproduced.
At N$^2$LO the points by Kievsky et al. are essentially exactly reproduced. This is a consequence of using a finite energy phase shift point to match the N$^2$LO three-body force.

\begin{figure}[H]
\center
\includegraphics[width=0.75\linewidth]{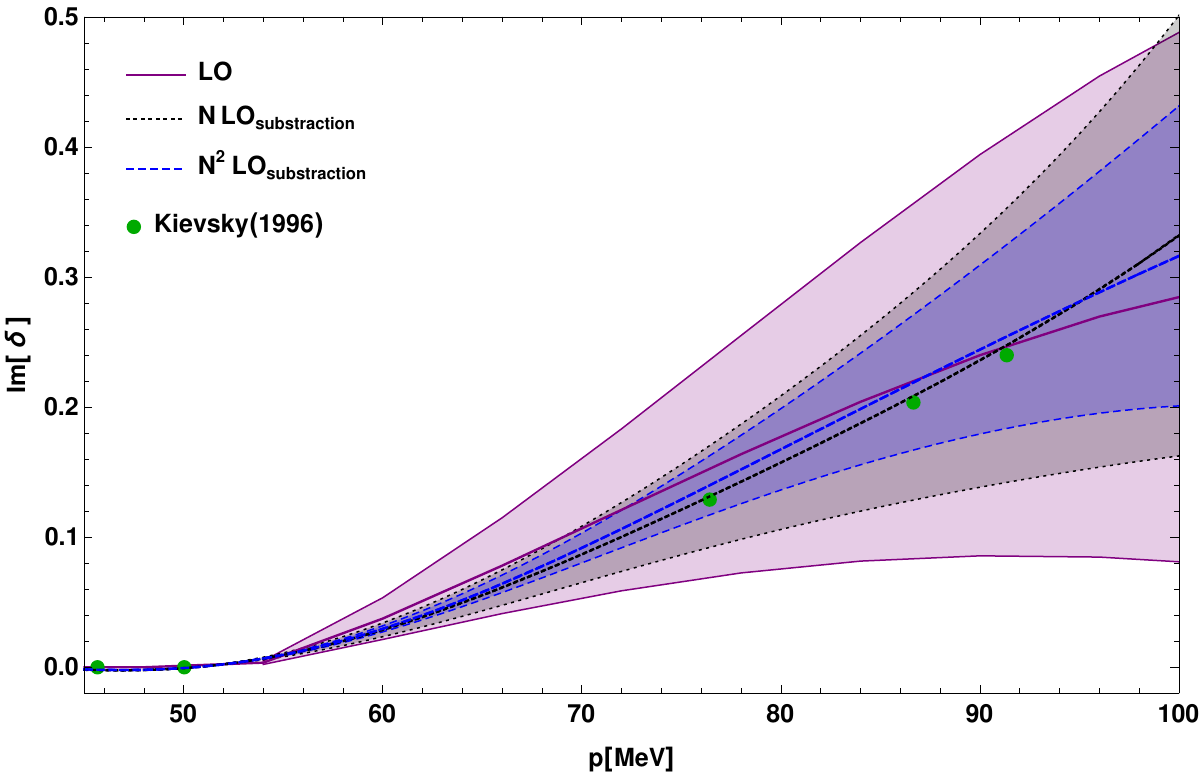}
\caption{The imaginary part of the neutron-deuteron $^2S_{1/2}$ phase shift in our alternative subtraction scheme. The purple solid line is the result at LO, while the black dotted 
blue dashed lines give our result at NLO and N$^2$LO. respectively.
The theoretical uncertainties is given by the shaded bands.
For comparison, we show he calculation 
by Kievsky et al.~\cite{Kievsky:1996ca} (circles).}
\label{PhaseshiftIm}
\end{figure}

The calculation of the imaginary part of the phase shift is shown 
Fig.~\ref{PhaseshiftIm}.
Here again the expected convergence behavior is observed. All three orders are consistent with each other and the 
calculation by Kievsky et al. \cite{Kievsky:1996ca} is well reproduced. 

In summary, we have shown that our alternative subtraction scheme for the non-perturbative inclusion of the effective range corrections can be applied in the $^2S_{1/2}$-channel of neutron-deuteron scattering. We reiterate that this resummation of range corrections does not extend the range of applicability of the EFT but is introduced as a matter of convenience to avoid divergences in finite volume applications.
However, as discussed in Sec. \ref{sec:su4} the justification for this scheme has to be modified. We {\em postulate} the property ${ W}(p,k,E)={K}(p,k,E)$ in the
subtracted Faddeev equations (\ref{eq:split}) (irrespective of the value of cutoff),
arguing that this identification merely amounts to a
resummation in the two-body sector, which is different from the standard form.
The resulting Faddeev equations lead to a consistent description of the neutron-deuteron system in   the $^2S_{1/2}$-channel. The next step is to test this scheme in finite volume calculations similar to the boson case discussed in Ref.~\cite{Pang:2022nim}.

\begin{acknowledgments}
 M.E. was supported by a PhD fellowship from Helmholtz Forschungsakademie Hessen f\"ur FAIR (HFHF).
  M.E. and H.-W.H. were supported by Deutsche Forschungsgemeinschaft
  (DFG, German Research Foundation) --
  Project ID 279384907 -- SFB 1245.
A.R.
was  supported in part by the Deutsche Forschungsgemeinschaft
(DFG, German Research Foundation) -- Project-ID 196253076 -- TRR 110,
Volkswagenstiftung 
(grant no. 93562) and the Chinese Academy of Sciences (CAS) President's
International Fellowship Initiative (PIFI) (grant no. 2024VMB0001).
\end{acknowledgments}


\begin{thebibliography}{99}

\bibitem{vanKolck:1997ut}
  U.~van Kolck,
  Lect.\ Notes Phys.\ {\bf 513}, 62 (1998).

\bibitem{Kaplan:1998tg}
  D.~B.~Kaplan, M.~J.~Savage and M.~B.~Wise,
  Phys.\ Lett.\ B {\bf 424}, 390 (1998).

\bibitem{Kaplan:1998we}
  D.~B.~Kaplan, M.~J.~Savage and M.~B.~Wise,
  Nucl.\ Phys.\ B {\bf 534}, 329 (1998).

\bibitem{vanKolck:1998bw}
  U.~van~Kolck,
  Nucl.\ Phys.\ A {\bf 645}, 273 (1999).

\bibitem{Chen:1999tn}
  J.-W.~Chen, G.~Rupak and M.~J.~Savage,
  Nucl.\ Phys.\ A {\bf 653}, 386 (1999).

\bibitem{Bedaque:2002mn}
P.~F.~Bedaque and U.~van Kolck,
Ann. Rev. Nucl. Part. Sci. \textbf{52}, 339 (2002).
  
\bibitem{Epelbaum:2008ga}
E.~Epelbaum, H.-W.~Hammer and U.-G.~Mei{\ss}ner,
Rev. Mod. Phys. \textbf{81}, 1773 (2009).

\bibitem{Hammer:2019poc}
H.-W.~Hammer, S.~K\"onig and U.~van Kolck,
Rev. Mod. Phys. \textbf{92}, 025004 (2020).

\bibitem{Hansen:2014eka}
M.~T. Hansen and S.~R. Sharpe,
Phys. Rev. D \textbf{90}, 116003 (2014).

\bibitem{Hansen:2015zga}
M.~T. Hansen and S.~R. Sharpe,
Phys. Rev. D \textbf{92}, 114509 (2015).

\bibitem{Hammer:2017uqm}
H.-W. Hammer, J.-Y. Pang, and A.~Rusetsky,
JHEP \textbf{09}, 109 (2017).

\bibitem{Hammer:2017kms}
H.~W. Hammer, J.~Y. Pang, and A.~Rusetsky,
JHEP \textbf{10}, 115 (2017).

\bibitem{Mai:2017bge}
M.~Mai and M.~{D\"{o}ring},
Eur. Phys. J. A \textbf{53}, 240 (2017).

\bibitem{Mai:2018djl}
M. Mai and M. D{\"{o}}ring,
Phys. Rev. Lett. \textbf{122}, 062503 (2019).

\bibitem{Muller:2021uur}
F.~M\"uller, J.~Y.~Pang, A.~Rusetsky and J.~J.~Wu,
JHEP \textbf{02}, 158 (2022).

\bibitem{Hansen:2019nir}
M.~T. Hansen and S.~R. Sharpe,
Ann. Rev. Nucl. Part. Sci. \textbf{69},  65 (2019).

\bibitem{Mai:2021lwb}
M. Mai, M. D\"oring, and A. Rusetsky,
Eur. Phys. J. ST \textbf{230}, 1623 (2021).

\bibitem{Hammer:2001gh}
H.-W.~Hammer and T.~Mehen,
Phys. Lett. B \textbf{516}, 353 (2001).

\bibitem{Bedaque:1998km}
P.~F.~Bedaque, H.-W.~Hammer and U.~van Kolck,
Nucl. Phys. A \textbf{646}, 444 (1999).

\bibitem{Ji:2011qg}
C.~Ji, D.~R.~Phillips and L.~Platter,
Annals Phys. \textbf{327}, 1803 (2012).


\bibitem{Ji:2012nj}
C.~Ji and D.~R.~Phillips,
Few Body Syst. \textbf{54}, 2317 (2013).

\bibitem{Vanasse:2013sda}
J.~Vanasse,
Phys. Rev. C \textbf{88}, 044001 (2013).

\bibitem{Bedaque:1997qi}
P.~F.~Bedaque and U.~van Kolck,
Phys. Lett. B \textbf{428}, 221 (1998).

\bibitem{Bedaque:1998mb}
P.~F.~Bedaque, H.~W.~Hammer and U.~van Kolck,
Phys. Rev. C \textbf{58}, R641-R644 (1998).


\bibitem{Ebert:2021epn}
M.~Ebert, H.~W.~Hammer and A.~Rusetsky,
Eur. Phys. J. A \textbf{57}, 332 (2021).

\bibitem{Pang:2022nim}
J.~Y.~Pang, M.~Ebert, H.~W.~Hammer, F.~M\"uller, A.~Rusetsky and J.~J.~Wu,
JHEP \textbf{07}, 019 (2022).

\bibitem{Bedaque:2002yg}
P.~F.~Bedaque, G.~Rupak, H.~W.~Griesshammer and H.~W.~Hammer,
Nucl. Phys. A \textbf{714}, 589 (2003). 

\bibitem{Dilg:1971gqb}
W.~Dilg, L.~Koester and W.~Nistler,
Phys. Lett. B \textbf{36}, 208 (1971).


\bibitem{Kievsky:1996ca}
A.~Kievsky, S.~Rosati, W.~Tornow and M.~Viviani,
Nucl. Phys. A \textbf{607}, 402 (1996).

\bibitem{vanOers:1967lny}
W.~T.~H.~van Oers and J.~D.~Seagrave,
Phys. Lett. B \textbf{24}, 562 (1967).

\end{thebibliography}
\end{document}